\documentclass[runningheads,a4paper]{llncs}

\usepackage{latexsym,bm}
\usepackage{amssymb}
\usepackage{graphicx}
\usepackage{bbm}
\usepackage{amsmath}
\usepackage{amssymb}
\usepackage{algorithmic}
\usepackage[linesnumbered,ruled,vlined]{algorithm2e}
\usepackage{url}
\usepackage{multirow}
\usepackage{array}

\urldef{\mails}\path|{jtsang, zhanghuaiwen2016, csxu}@nlpr.ia.ac.cn|
\begin{document}

\mainmatter  % start of an individual contribution

\title{Visual BFI: an Exploratory Study for Image-based Personality Test}

%\titlerunning{Lecture Notes in Computer Science: Authors' Instructions}

\author{Jitao Sang, Huaiwen Zhang, Changsheng Xu}

\institute{Institute of Automation, Chinese Academy of Sciences,\\100190 Beijing, China\\
\mails
%\path|{jtsang}@nlpr.ia.ac.cn|
}
%\url{http://www.springer.com/lncs}}
%
%%
%% NB: a more complex sample for affiliations and the mapping to the
%% corresponding authors can be found in the file "llncs.dem"
%% (search for the string "\mainmatter" where a contribution starts).
%% "llncs.dem" accompanies the document class "llncs.cls".
%%
%
%\toctitle{Lecture Notes in Computer Science}
%\tocauthor{Authors' Instructions}

\maketitle

\begin{abstract} This paper positions and explores the topic of image-based personality test. Instead of responding to text-based questions, the subjects will be provided a set of ``choose-your-favorite-image'' visual questions. With the image options of each question belonging to the same concept, the subjects' personality traits are estimated by observing their preferences of images under several unique concepts. The solution to design such an image-based personality test consists of concept-question identification and image-option selection. We have presented a preliminary framework to regularize these two steps in this exploratory study. A demo version of the designed image-based personality test is available at \url{http://www.visualbfi.org/}. Subjective as well as objective evaluations have demonstrated the feasibility of image-based personality test in limited questions.
\end{abstract}
% this paper presents solution to..
% in the form of select-your-favorite-image,

% introduce a gradient boosted solution, explore the representative concepts for question, and images for options
% evaluation on prediction and real-world

\section{Introduction}
Personality refers to a type of psychological traits explaining human behaviors in terms of a few, stable and measurable individual characteristics~\cite{matthews2003personality}. Different from demographic attributes, personality traits explain and predict behavior differences from the internal psychological perspective. One of the most popular personality models is Big Five (BF) or Five-Factor Model (FFM)~\cite{mccrae1992introduction}, which defines personality along five dimensions, i.e., \emph{Openness}, \emph{Conscientiousness}, \emph{Extraversion}, \emph{Agreeableness} and \emph{Neuroticism}. Accurately estimating these personality traits has wide applications including occupational assistance~\cite{barrick1991big}, target advertisement~\cite{chittaranjan2011s}, personalized recommender system~\cite{tkalcic2009emotive}, disease detection and prevention~\cite{clarkin1996treatment}, and even human-robot interaction~\cite{walters2005influence}.

\begin{figure}[t]
\centering
\includegraphics[width=0.99\textwidth]{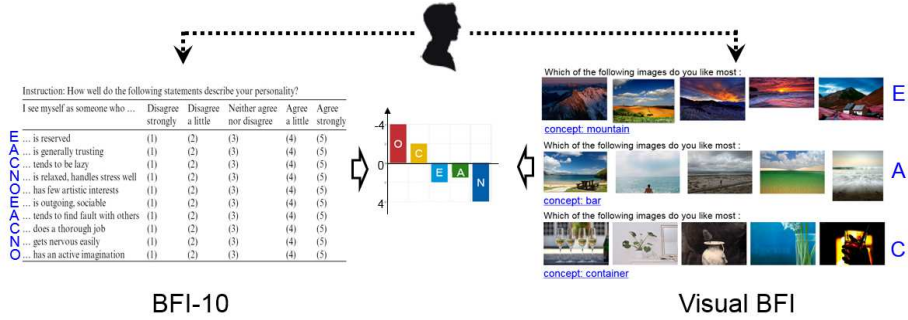}
\vspace{-4mm}
\caption{Personality test: text-based BFI-10 v.s. visual BFI.}
\label{fig1}
\vspace{-5mm}
\end{figure}

Traditionally, standard personality tests, e.g., Big-Five Inventory (BFI), are used to gauge one's personality score in each dimension. In these tests, psychological experts design text-based questionnaires to ask subjects make self-assessment by responding the level of agreement to each question. The left of Fig.~\ref{fig1} illustrates a shorter version of the BFI comprising of 10 questions (i.e., BFI-10). The suffix ``O, C, E, A, N'' is the abbreviation of five personality dimensions, indicating which dimension the question contributes to. Despite the wide utilization of BFI, the current text-based personality tests are subject to several limitations: (1) The subjects need to read and understand each question thoroughly before making responses. This is time consuming and can be a huge burden to the subjects especially in long personality tests~\cite{paulhus2007self} (e.g., BFI-44 with 44 questions, NEO-PI-R with 240 questions). Moreover, the text-based personality tests are built on the assumption that the subjects ``have access to the psychological property to be measured'' and know enough about themselves to make accurate response~\cite{mcdonald2008measuring}. This is impractical in many cases and significantly limits the scope of applicability. (2) The text-based questions convey clear meanings to portray the subjects. With the predisposition toward self-enhancement, the subjects are likely to respond in a way presenting themselves more favorable~\cite{fiske1991social}. For example, regarding the question ``I see myself as someone who tends to be lazy'' in BFI-10, most subjects will have ``disagree'' response to maintain positivity about themselves even at the expense of being unrealistic. This easily leads to response biases and inaccurate personality estimation. (3) Text-based questionnaires are language-sensitive. Language-specific models need to be carefully developed by experts and professionals, instead of just translating a reference model into a destination language~\cite{de1998lingua}. For example, ``calm'' is used to measure \emph{Neuroticism} in English-based BFI. The direct translation of ``calm'' in German is ``ruhlg''. However, ``ruhlg'' in German actually has both correlation with \emph{Neuroticism} and \emph{Extraversion}.

Recently has witnessed some studies attempting to automatically infer personality traits from users' social media interactions with images. Cristani et al. proposed to assess the personality of users by looking at their favorite images~\cite{cristani2013unveiling}. 300 Flickr users are examined with each consisting of 200 favorite images. Following this, in~\cite{guntuku2015personality}, more image features are extracted for personality prediction. Based on the derived personality traits, the application of image recommendation is investigated. These attempts suggest that people's image favorite behavior promisingly reflects their personality traits. Inspired by this, to address the above-mentioned problems in text-based personality tests, we propose to research on image-based personality test, by exploring the underlying correlation between subjects' visual preferences and personality traits to select and organize discriminative images for questionnaire design.

Language psychology shows that the choice of words is driven not only by meaning, but also by speakers'/writers' psychological characteristics such as emotions and personality traits~\cite{tausczik2010psychological}. In other words, when expressing the same meaning, it is highly possible the different psychological characteristics that leads to different word choices. We are motivated to make an analogy in the scenario of image choice, and predict the subjects' personality traits by investigating their preferences of images belonging to the same concept. In particular, \emph{the goal is to design a set of ``choose-your-favorite-image'' questions, with each question corresponding to one concept and options for each question corresponding to different patterns of images under this concept}. On the right of Fig.~\ref{fig1} we show three example questions in our designed visual BFI. This visual BFI preferably solves the three problems in text-based personality tests: (1) Image is recognized as more natural interaction means than text. With less sense of task-performing, subjects are expected to answer the questionnaire in a more relaxed way. (2) The intent behind choosing images is not clear, which is less offensive to the subjects. Subjects therefore make objective responses based on their realistic perceptions. (3) People's perception to visual information is universal regardless of their mother tongue. Aside from cultural differences, we hope the image-based personality test is applicable to subjects in different languages.

Designing the image-based personality test consists of concept-question identification for each personality dimension, and image-option selection for each question. In the remainder of this paper, Section 2 explores the potential of individual concept in personality prediction. In Section 3, we introduce how to combine several concepts and develop boosted regressor in predicting each dimension of personality, and how to select images from these concepts to construct questions and options for questionnaire design. Section 4 presents experimental results of both the proposed boosted regressor in automatical personality prediction, and the designed image-based questionnaire on real-world personality test via Amazon Mechanical Turk.

%我们的思路

%对应说好处：
%%（2）language-independent: circumvent this problem, aside from the cultural difference, human's perception to visual is universal --(metaphors in German and Japanese are presented as evidence for the existence of a language-independent mechanism responsible for metaphor production)
%%--Unlike text, an image is a universal representation; we can easily understand the semantic content of images regardless of our mother tongue.

\section{Exploring Single Concept for Personality Prediction}
\begin{figure}[t]
\begin{minipage}[b]{0.405\linewidth}
\centering
\includegraphics[width=0.97\textwidth]{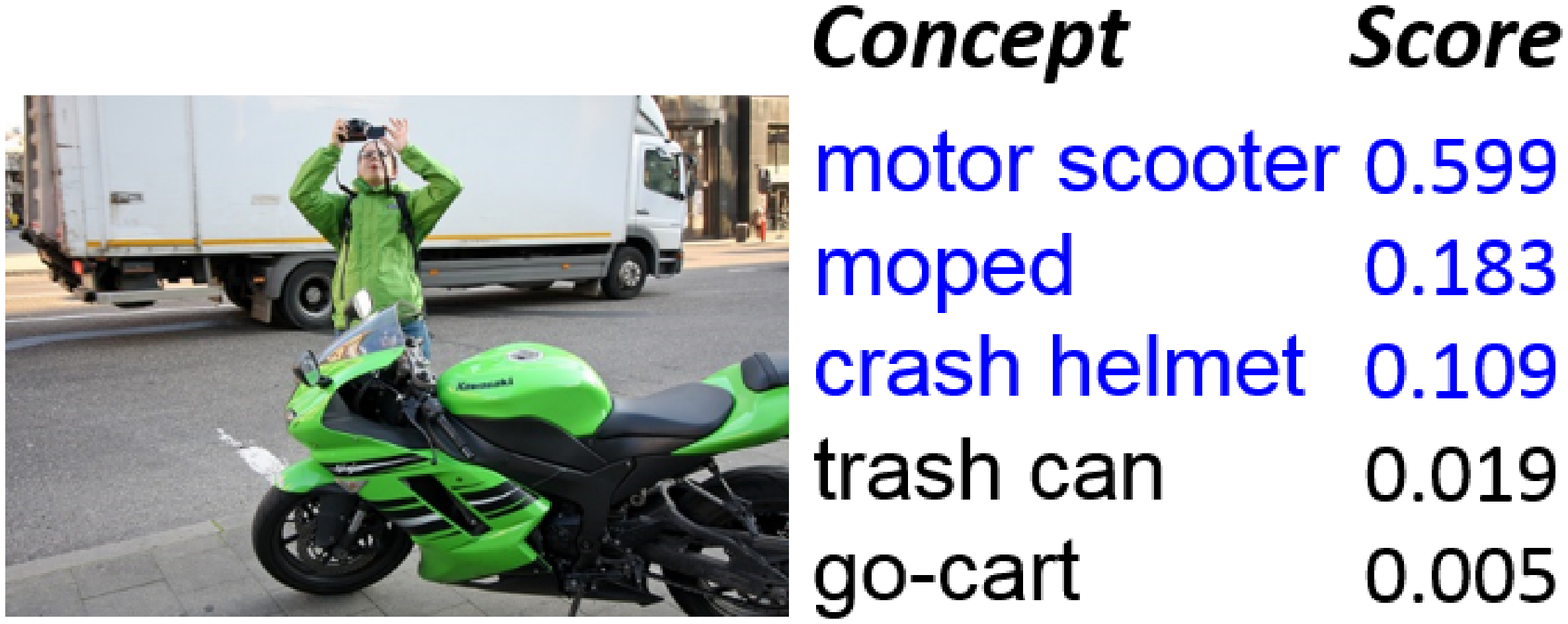}
\centerline{(a)}
\end{minipage}
\begin{minipage}[b]{0.594\linewidth}
\centering
\includegraphics[width=0.98\textwidth]{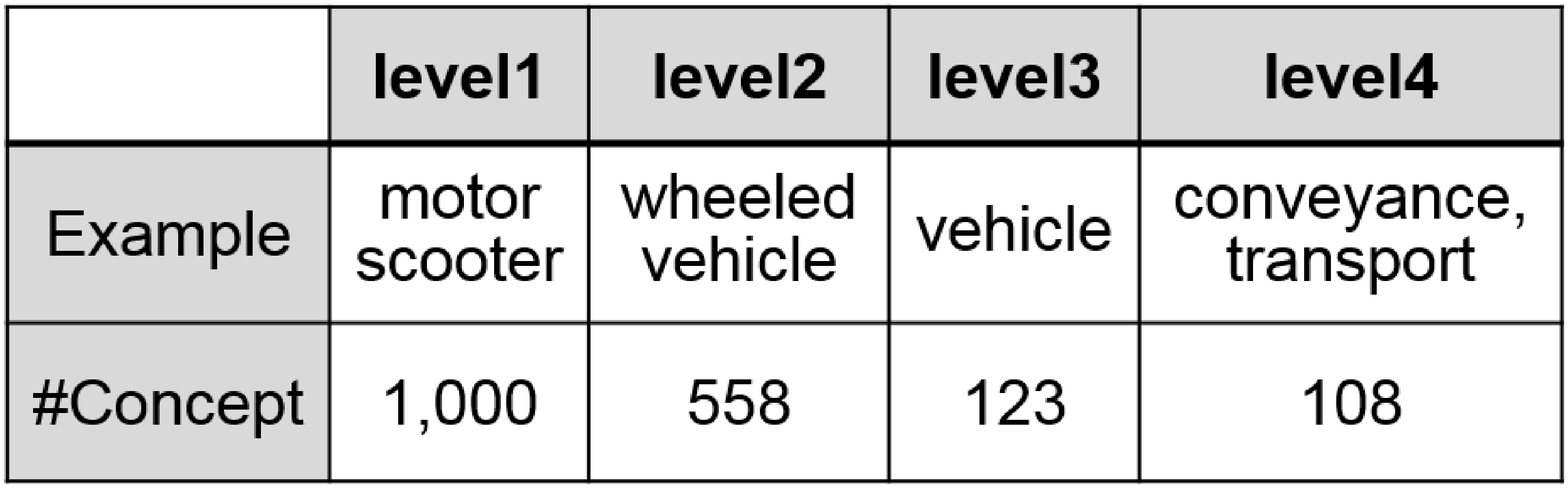}
\centerline{(b)}
\end{minipage}\vspace{-2mm}
\caption{Candidate concept generation: (a) GoogleNet-based concept detection (the remained concepts are highlighted with blue); (b) the example and statistic for hypernym-based concept expansion.}
 \label{fig:2}
 \end{figure}

%In this section, we first introduce the derivation of concept and then present the personality prediction performance based on subjects' responses to images belonging to individual concepts.
Our study is based on the PsychoFlickr dataset provided in~\cite{cristani2013unveiling}. This dataset consists of 300 Flickr users, with each user associating with his/her 200 favorite images and the self-assessed personality traits in five dimensions. 82-dimension aesthetics and content feature has been extracted for each image. The first step of the study is to generate a candidate concept list. As illustrated in Fig.~\ref{fig:2}(a), for each of the 60,000 images, GoogleNet~\cite{szegedy2015going} was used to obtain the confidence score over 1,000 ImageNet categories and the top-5 categories with confidence score larger than 0.1 are remained. The 1,000 ImageNet categories construct our original concept set at level 1. To expand the candidate concepts, for each of 1,000 concepts at level 1, its hypernym in WordNet is traced and added into the included concepts for corresponding images. We repeat this process three times to obtain totally 1,789 concepts at four levels. Fig.~\ref{fig:2}(b) shows an example concept hierarchy and the number of traced unique concepts at each level.

 \begin{table}[t]
\caption{Personality prediction accuracy in terms of RMSE.}
\label{tab:1}
\vspace{-6mm}
\begin{center}
\begin{tabular}{|c|c|c|c|c|c|c|c|c|c|c|}
\hline
{\textbf{Trait}}&\multicolumn{3}{c|}{\textbf{SVR}}& \multicolumn{3}{c|}{\textbf{LASSO}} &  \multicolumn{3}{c|}{\textbf{CART}} & {\textbf{CG+LASSO}~\cite{cristani2013unveiling}}     \\\hline
{\textbf{O}}&{1.624}&{1.639}&{1.656}&{\textbf{1.612}}&{1.622}&{1.638}&{1.621}&{1.625}&{1.630}&{1.698}\\\hline
{\textbf{C}}&{\textbf{1.729}}&{1.745}&{1.768}&{1.730}&{1.742}&{1.749}&{1.808}&{1.813}&{1.817}&{1.789}\\\hline
{\textbf{E}}&{2.124}&{2.154}&{2.166}&{2.111}&{2.121}&{2.133}&{2.106}&{2.182}&{2.228}&{\textbf{2.077}}\\\hline
{\textbf{A}}&{1.612}&{1.639}&{1.653}&{1.610}&{1.617}&{1.632}&{\textbf{1.606}}&{1.614}&{1.631}&{1.669}\\\hline
{\textbf{N}}&{\textbf{2.148}}&{2.160}&{2.172}&{2.181}&{2.207}&{2.225}&{2.183}&{2.239}&{2.314}&{2.208}\\\hline
\end{tabular}
\end{center}\vspace{-4mm}
\end{table}

The images including concept $c$ construct a image set $\mathcal{I}_c$, and the users who has favored image $I_i\in\mathcal{I}_c$ constructs a user set $\mathcal{U}_c$. Our goal in this section is to examine the potential of personality prediction based on users' favorite images belonging to single concept. To this goal, among the 1,789 concepts, we first identified 235 concepts that are favored by at least 104 users~\footnote{~\small{For those concepts favored by more than 104 users, 104 random users are selected to construct the sample set for this concept. We fix the sample number as 104 to facilitate the performance comparison with solutions in the next section.}}, i.e., $\forall c\in\mathcal{C}_1, |\mathcal{U}_c|\geq 104$. For one concept $c\in\mathcal{C}_1$, we assume each user $u\in\mathcal{U}_c$ only have one favorite image and utilize the 82-dimension image feature as user representation $\mathbf{x}_u^{(c)}$. Different personality traits are treated separately, and for each personality trait $p\in\{O,C,E,A,N\}$, we need to build a set of concept-based regressors: 104 user samples for each concept, with each sample's input as the user's favorite image feature vector $\mathbf{x}_u^{(c)}$, and the output as the user's personality score $p_u$ (integer value from $-4$ to $4$).

Three standard regression methods are utilized: Support Vector Regression (SVR), LASSO regression, Classification And Regression Tree (CART). Considering the small scale of samples, 10 times of 10-fold cross validation is conducted and only statistically significant (\emph{p-value}$<5\%$) results are reported. Table~\ref{tab:1} shows the prediction accuracy of top-3 single concepts for each method. For comparison, the performance based on the method proposed in~\cite{cristani2013unveiling} is also shown and denoted as CG+LASSO. Note that the study in~\cite{cristani2013unveiling} considered 200 favorite images for each user. While, the results reported under {SVR}, {LASSO} and {CART} only considered at most 5 favorite images for each user. It is demonstrated from the results that, by examining users' favorite images belonging to few selective concepts, we can achieve comparable, if not better prediction accuracy than that based on much more unorganized favorite images.

\begin{center}
\begin{algorithm}[t]
\caption{View-based Gradient Boosted Decision Tree (vGBDT)}
\label{alg:1}
\KwIn{User's personality score at certain trait $p_i\in[-4,4]$, and his/her $K$-view representation $\mathbf{x}_i\in \mathbb{R}^{Kd}$, $i=1,\cdots,N$ ($N$ is the number of training samples). }

\KwOut{The strong regressor $F: \mathbf{x}\rightarrow p$.}
\vspace{1mm}
$F_0=\frac{1}{N}\sum_{i=1}^Np_i$\\
\For{$ m=1$ to $M$}
  {
        $r_i = p_i-F_{m-1}(\mathbf{x}_i), i=1,\cdots,N$\\
        $(V_m, R_m, A_m)=\arg \underset{V,R,A}\min\sum_{i=1}^N||r_i-T(\mathbf{x}_i;V,R,A)||_2^2$\\
        $F_m(\mathbf{x})=F_{m-1}(\mathbf{x})+\upsilon T(\mathbf{x};V_m,R_m,A_m)$
  }

\textbf{Return}: $F=F_M$.
\end{algorithm}
\end{center}
\vspace{-17mm}

\section{Combining Multiple Concepts for Personality Test}
The previous section has proved the feasibility of automatical personality prediction by only examining user's image favorite behavior under single concept. This section further this study by introducing: (1) how to combine users' image favorite behavior under multiple concepts to improve the prediction accuracy, and (2) how to exploit the developed prediction model for image-based personality test design.

\subsection{vGBDT-based Multiple Concept Combination}
A natural way to combine different base models is ensemble learning. Considering totally $K$ concepts, each user can be seen as a $K$-view sample represented as $\mathbf{x}=(\mathbf{x}^{(1)},\cdots,\mathbf{x}^{(i)},\cdots,\mathbf{x}^{(K)}]^\mathrm{T}$, where $\mathbf{x}^{(i)}=[x_{i(d-1)+1},\cdots,x_{id}]$ indicates his/her favorite image features under the $i^{th}$ concept. To facilitate the questionnaire design, each base regressor is expected to correspond to one single concept, so as to collect user's image favorite response for one concept in each round of question. This means only one view in user's representation will be utilized in one base regressor and then contribute to the final personality trait estimation.

\begin{figure}[t]
\centering
\includegraphics[width=0.95\textwidth]{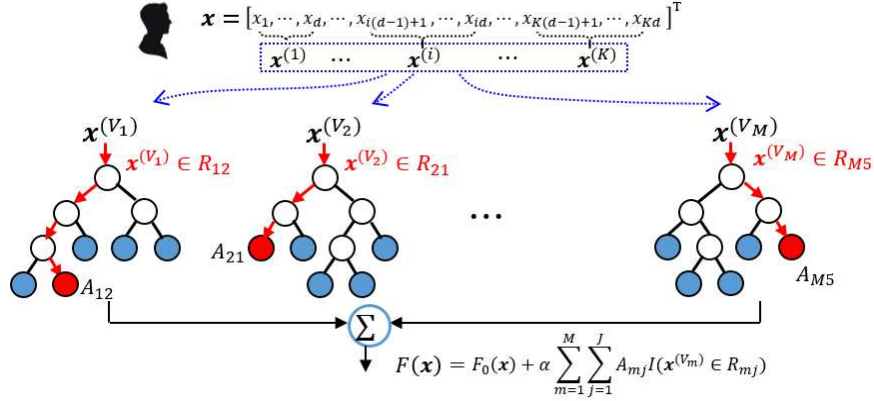}
\vspace{-18mm}
\caption{vGBDT-based personality prediction by combining multiple concepts.}
\label{fig:3}
\vspace{-4mm}
\end{figure}

Standard ensemble learning methods cannot be directly applied in this scenario. For example, Gradient Boosted Decision Tree (GBDT) select the best base CART from a unique feature space to fit to the residual. In this study, we modify the standard GBDT and introduce a view-based GBDT (vGBDT) to address this problem. Specifically, in the training phase, for each round of base regressor, vGBDT not only tune the optimal partitions and the output leaf value, but identify which view of features will be utilized. The base regression tree is denoted as $T(\mathbf{x};V_m,R_m,A_m)=\sum_{j=1}^JA_{mj}\cdot\mathbb{I}(\mathbf{x}^{(V_m)}\in R_{mj})$, where $\mathbb{I}(\cdot)$ is the indicator function, $J$ is the number of leaf nodes, $V_m, R_m, A_m$ indicates the selected view of feature, learned disjoint partitions, and output leaf values respectively. The training phase of vGBDT is summarized in Algorithm~\ref{alg:1}. Assuming $M$ base regressors (concepts) are considered, the final strong regressor can be expressed as an ensemble of regression trees $T$:\\\vspace{-2.5mm}
\begin{equation}
\label{eq:1}
F(\mathbf{x};\mathbf{\Theta})=F_0+\upsilon\sum_{m=1}^{M}T(\mathbf{x};V_m,R_m,A_m)
\end{equation}
where $F_0$ is the mean value of all training samples, $\upsilon$ is the shrinkage parameter. In the testing phase, as illustrated in Fig.~\ref{fig:3}, user's $V_m^{th}$ view feature is used as input to the $m^{th}$ regressor and the final personality score is calculated according to Eqn.~\eqref{eq:1}.

%However,
%
%
%- a natural way for combinbing is 提升方法
%- however, standard提升方法 is based on the same feature representation.
%- we modify it and present a view-based... to address...
%
%- multi-view symble

\subsection{Image-based Personality Test Design}
After training vGBDT for each personality trait, $M$ concepts are identified for the $M$ corresponding questions in the personality questionnaire. For each concept-question, we need then to select $J$ representative images as options for subjects to choose from.

\begin{figure}[t]
\centering
\includegraphics[width=0.92\textwidth]{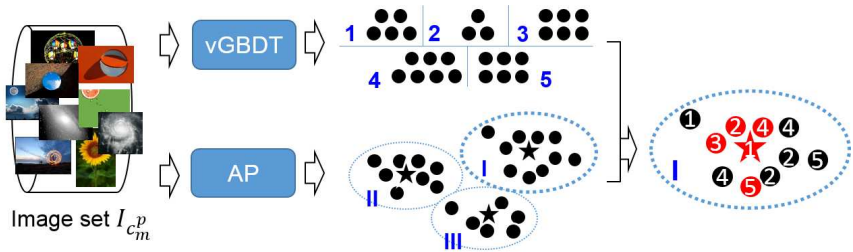}
\vspace{-3mm}
\caption{Illustration for image-option selection (the red ones are selected).}
\label{fig:4}
\vspace{-4mm}
\end{figure}

Assuming the $m^{th}$ concept in predicting personality trait $p$ is $c_m^p$, each image $I_i\in\mathcal{I}_{c_m^p}$ can be assigned to one of the $J$ leaf nodes by running the base regressor $T(V_m^p,R_m^p,A_m^p)$. This is illustrated in the top of Fig.~\ref{fig:4} ($J=5$ in this case, where each image is assigned a label $l_i\in[1,5]$). To reduce the distractors that influence subjects' choice, it is critical to make the $J$ image-options as similar as possible, e.g., at the similar aesthetic level, all w/ or w/o faces in the images, etc. Therefore, we conduct Affinity Propagation (AP) on the image set $I_i\in\mathcal{I}_{c_m^p}$ to obtain several image clusters. Within the largest image cluster (cluster \textrm{I})~\footnote{~\small{In practical implementation, we can select images from different clusters to design several versions of questionnaires.}}, for each label from $\{1,2,\cdots,J\}$, the image nearest to the cluster center is selected as the option image (see the bottom of Fig.~\ref{fig:4}).

During the personality test, for each personality trait $p$, the subjects will be asked to sequentially answer $M$ questions, with each question consisting of $J$ image-options to choose from. The subject's final personality score for $p$ is calculated by collecting his/her choices:\\\vspace{-2.5mm}
\begin{equation}
F=F_0+\upsilon\sum_{m=1}^MA_{m\pi_m}
\end{equation}
where $\pi_m$ records the option index of subject's choice for the $m^{th}$ question. An online version of the designed image-based personality test is available at \url{http://www.visualbfi.org/}. The personality test is expected to be finished within 5 minutes.

\section{Experiments}
\subsection{Evaluation on Automatic Personality Prediction}
To evaluate the performance of combining multiple concepts for personality prediction, among the 1,789 concepts, we identified 36 concepts that are co-favored by 104 users to construct he candidate concept set $\mathcal{C}_2$, i.e., $|\bigcap_{c\in\mathcal{C}_2}\mathcal{U}_c|= 104$. With the view number $K=36$, the task of vGBDT is to select and construct $M$ concept-based base regressors by examining the training users' favorite images over 36 concepts.

For parameter setting, considering the practical application in personality test, we choose a small number of base regressors (concept-questions) $M=5$, i.e., the designed personality test will totally consist of $5\times 5=25$ concept-questions. For each question, the number of image-options is chosen as $J=5$. A large shrinkage weight is selected as $\upsilon=0.5$ to value the contribution of each question. 10 times of 10-fold cross validation are conducted, with the average personality prediction accuracy reported in Table~\ref{tab:2}. \emph{Single Concept} indicates the best performance of single content for each personality trait in Table~\ref{tab:1}. It is shown that by combining users' image favorite responses over multiple concepts, vGBDT achieves significant performance gains.

%~\footnote{~\small{Note that some studies use the range of personality score as [1,5] instead of [-4,4]. Their reported RMSE can be lower than $1.0$~\cite{quercia2011our,guntuku2015personality}.}}

\begin{table}[t]
\caption{Personality prediction accuracy in terms of RMSE.}
\label{tab:2}
\vspace{-5mm}
\begin{center}
\begin{tabular}{|c|c|c|c|}
\hline
{\textbf{Trait}}& {\textbf{Single Concept}} & {\textbf{CG+LASSO}~\cite{cristani2013unveiling}}& {\textbf{vGBDT}}     \\\hline
{\textbf{O}}&{1.612}&{1.698}&{\textbf{1.232}}\\\hline
{\textbf{C}}&{1.729}&{1.789}&{\textbf{1.571}}\\\hline
{\textbf{E}}&{2.106}&{2.077}&{\textbf{1.601}}\\\hline
{\textbf{A}}&{1.606}&{1.669}&{\textbf{1.248}}\\\hline
{\textbf{N}}&{2.148}&{2.314}&{\textbf{1.796}}\\\hline
\end{tabular}
\end{center}\vspace{-4mm}
\end{table}

\begin{figure}[t]
\begin{minipage}[b]{0.499\linewidth}
\centering
\includegraphics[width=0.98\textwidth]{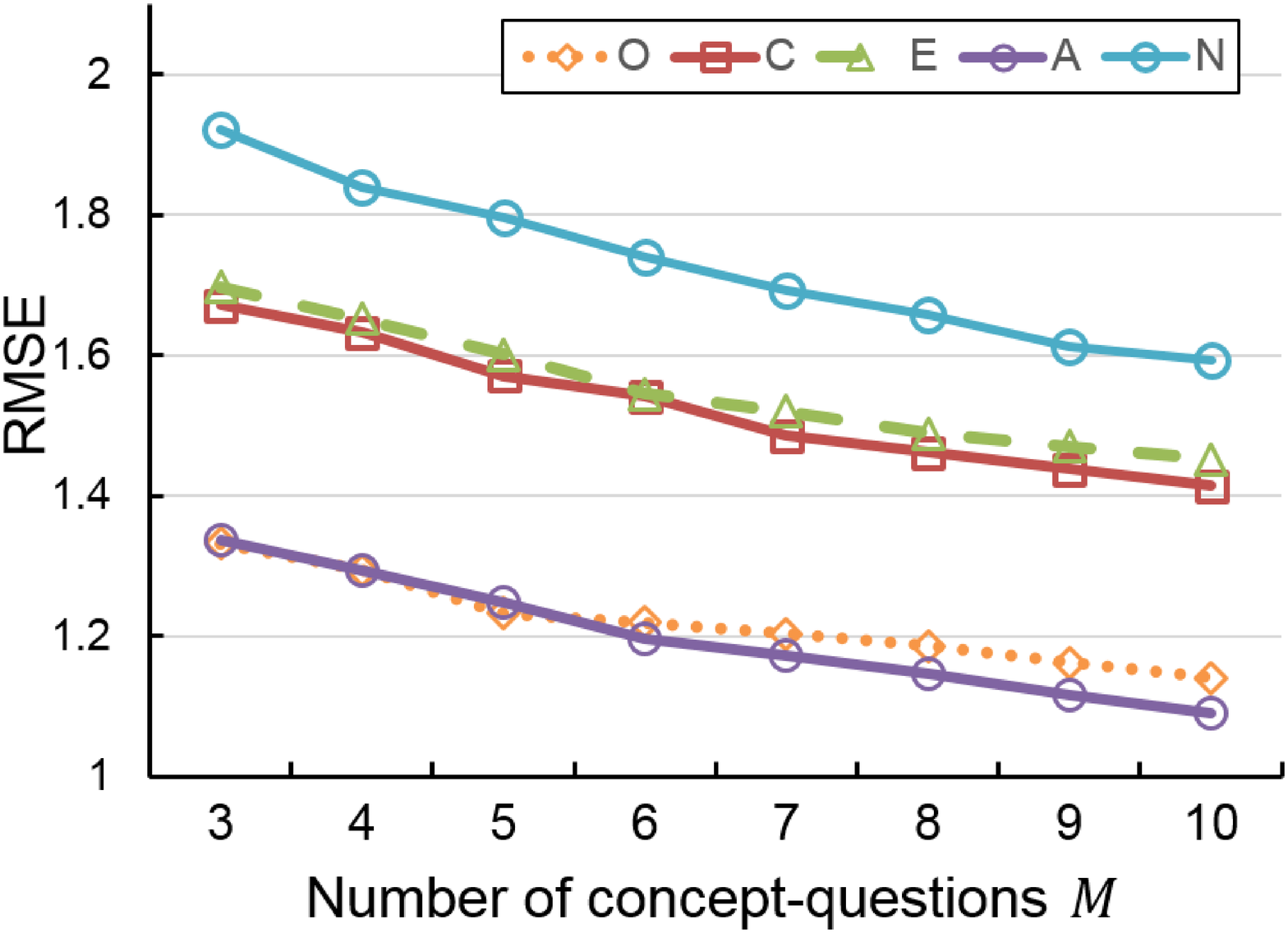}
\centerline{(a)}
\end{minipage}
\begin{minipage}[b]{0.499\linewidth}
\centering
\includegraphics[width=0.98\textwidth]{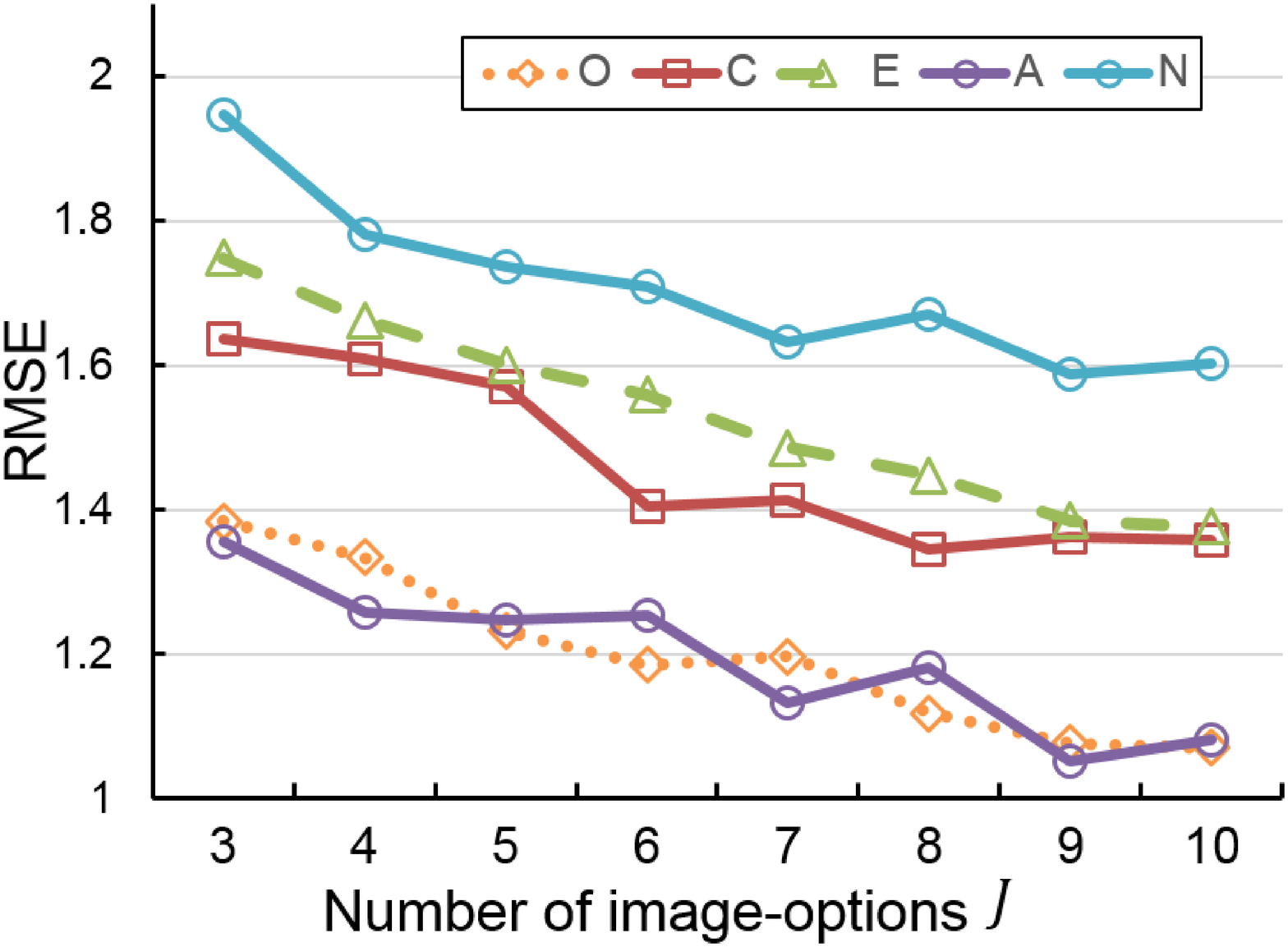}
\centerline{(b)}
\end{minipage}\vspace{-1.5mm}
\caption{Influence of vGBDT parameters: (a) the number of concept-questions $M$, (b) the number of image-options $J$.}
 \label{fig:5}\vspace{-4mm}
 \end{figure}

We also examined the influence of concept-question and image-option numbers on prediction performance. Fig.~\ref{fig:5}(a) shows the personality prediction RMSE with fixed image-option number $J=5$ and different concept-question numbers $M$ for each trait. The RMSE consistently decreases as the number of concept-questions increases, to achieve a low RMSE around 1.1 for the trait \emph{Agreeableness} and \emph{Openness} with $M=10$ concept-questions. This basically indicates that in the personality test, the more users' favorite choices over concepts are observed, the more accurate personality estimation results can be obtained. Fig.~\ref{fig:5}(b) shows the prediction RMSE with fixed concept-question number $M=5$ and different image-option numbers $J$. Different from that of Fig.~\ref{fig:5}(a), there exists no monotonically decreasing tendency in Fig.~\ref{fig:5}(b). For an acceptable personality test, we fix the number of image-options $J=5$ and the number of concept-questions $M=5$ in the questionnaire design and the later real-world evaluation.

\subsection{Evaluation on Real-world Personality Test}
We conducted a real-world evaluation by recruiting 67 master workers from Amazon Mechanical Turk (MTurk). Each subject was asked to answer four questionnaires: two versions of the proposed visual BFI (vBFI), one BFI-10 and one BFI-44. To guarantee the credibility of the responses, we examined the stableness between subjects' derived traits from BFI-10 and BFI-44. 40 MTurk subjects were kept for evaluation who had RMSE in BFI-10 v.s. BFI-44 lower than $1.2$.

We first compared between the personality traits derived from the first version of visual BFI (vBFI$\_1$) and BFI-10. The results are shown in the second column of Table~\ref{tab:3}. We see a higher RMSE between 1.5 to 2.0 than that in Table~\ref{tab:2}. We ascribe this decreased accuracy to two reasons: (1) different sample distribution between the training Flickr users and the testing MTurk workers (e.g., the different mean value); and (2) the coarser user division by locating subjects in fixed and limited image-options. It leaves us space for improvement by considering these two issues in the future work.

\begin{table}[t]
\caption{Real-world evaluation results from MTurk.}
\label{tab:3}
\vspace{-5mm}
\begin{center}
\begin{tabular}{|c|m{3.3cm}<{\centering}|m{1.8cm}<{\centering}|m{3.3cm}<{\centering}|}
\hline
{\textbf{Trait}} & {\textbf{vBFI$\_1$ v.s. BFI-10} (RMSE)} & {\textbf{Rate} (mean/std.)} & {\textbf{vBFI$\_1$ v.s. vBFI$\_2$} (RMSE)}     \\\hline
{\textbf{O}}&{1.647}&\multirow{5}{*}{5.150 /1.494)}&{0.872}\\\cline{1-2}\cline{4-4}
{\textbf{C}}&{1.859}&&{1.004}\\\cline{1-2}\cline{4-4}
{\textbf{E}}&{2.059}&&{1.101}\\\cline{1-2}\cline{4-4}
{\textbf{A}}&{1.506}&&{0.866}\\\cline{1-2}\cline{4-4}
{\textbf{N}}&{2.075}&&{1.300}\\\hline
\end{tabular}
\end{center}\vspace{-5mm}
\end{table}

Although text-based personality test like BFI-10 has recognized accuracy and is utilized as the ground-truth in our previous studies, it should be noted that the goal of image-based personality test is not to fit to the text-based test results, but to match with the subjects' own perception. Therefore, in addition to comparing with BFI-10, we also solicited the subjects' perception of the derived traits. After finishing vBFI$\_1$, we presented the estimated personality results from vBFI~\footnote{~\small{Note that the subjects were never given their personality test results from text-based BFI-10 or BFI-44 to avoid the interaction effect of different tests.}} to the subjects with the detailed explanation for each trait. Each subject rated how accurate they thought the derived traits were on a seven-likert scale (1 being worse and 7 being best). The mean of the resultant ratings is 5.150 ($std=1.494$), suggesting that the derived traits from vBFI generally matched well with their own perceptions.

Finally the robustness of the designed visual BFI is examined. Following the comparison between vBFI and BFI-10, we compared between the derived traits from the two versions of visual BFI (vBFI$\_1$ and vBFI$\_2$). Note that vBFI$\_2$ is designed by selecting image-options from the second-largest cluster as footnoted in Section 3.2. Results shown in the last column of Table~\ref{tab:3} achieve a relative low RMSE around $1$. This robustness demonstrates the feasibility of visual BFI in accurately personality estimation from another perspective.

\section{Conclusion, Discussion and Future Work}
In this paper we position the topic of image-based personality test design and present our first exploratory study under this topic. Subjective as well as objective evaluations have demonstrated the feasibility of personality estimation by observing subjects' ``choose-your-favorite-image'' responses over few concept-questions.

\begin{figure}[t]
\centering
\includegraphics[width=0.999\textwidth]{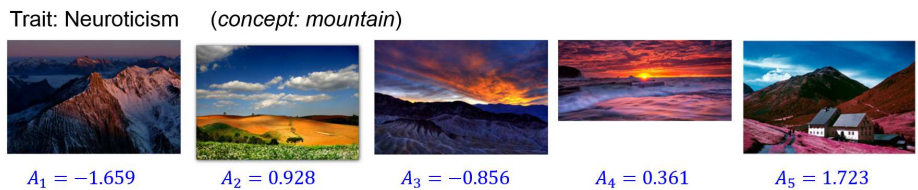}
\vspace{-5mm}
\caption{Example of image-options with the corresponding output leaf value $A$.}
\label{fig:6}
\vspace{-6mm}
\end{figure}

Under the topic of image-based personality test design, this work can be extended along several directions in the future: (1) The relative small scale of subjects limits the number of explored concepts in current study. We are working towards collecting the personality traits as well as their favorite image behaviors from more subjects, so as to explore more concepts for personality test design. (2) Regarding the task of personality estimation, the image feature should reflect the style of users who favorite it, instead of indicating the image semantics. Since it is difficult to pre-define what features contribute to the discrimination of user personality, the second future direction is to combine deep learning to extract and discover the most contributive features. (3) The current personality test is based on a static questionnaire. An interesting extension is to formulate the personality test as a dynamic decision making process and develop solutions for more efficient dynamic questionnaires: given subject's previous responses, to dynamically select subsequent questions with goal of accurate personality estimation in as few steps as possible. (4) It is also significant to investigate into the mechanism behind the correlation between the visual preference behaviors and personality traits from a psychological perspective. Taking Fig.~\ref{fig:6} as example: why different image favorites on the concept ``mountain'' discriminates the trait of \emph{Extraversion}, and why favoring certain style of ``mountain'' images contribute most to the score calculation (with most positive output leaf value). (5) We realized that personality is only one of the factors leading to the visual preference difference. Therefore, a critical future direction is to exclude the other influencing factors like age, gender and cultural background, by restricting the subjects' to fall into certain group. The ideal solution is to develop a unified visual personality test model, and optimize model parameter settings for different groups of subjects.

In addition to image-based personality test design, this work is potentially interest-provoking to two closely-related problems: (1) Data-driven questionnaire design. Questionnaire design has long been viewed as more of an art than a science~\cite{rattray2007essential}, where questions are initialized by professionals and then modified according to the collected responses from surveys. In the era of big data, as demonstrated in this study, it has great potential in exploring correlations from historic data to shed light on question development, or completely automating questionnaire design by formulating an optimization problem. (2) Active user modeling. Users are basically willing to answer few questions if more accurate user profiles and improved personalized services are promised~\cite{birlutiu2013efficiently}. Therefore, actively collecting users' responses to carefully-designed questions can serve as important supplements to traditional user modeling based on numbers of noisy and low-quality passive data. Similar to the personality test problem, the key in active user modeling is the trade-off between efficiency and effectiveness: to achieve an acceptable user modeling accuracy in minimum questions before annoying the user.

\section{Acknowledgement}
The authors thank Cristina Segalin for providing the PsychoFlickr dataset and the code of \emph{CG+LASSO}.

%\bibliographystyle{abbrv}
%\small
%\bibliography{videoRec}

\scriptsize
\bibliographystyle{unsrt}
\bibliography{sigproc}

\end{document}